\documentclass[twocolumn,showpacs,preprintnumbers,amsmath,amssymb,prb,floatfix]{revtex4}

\usepackage{graphicx}
\usepackage{dcolumn}
\usepackage{bm}
\usepackage[latin1]{inputenc}

\newcommand{\bra}{\langle}
\newcommand{\ket}{\rangle}

\begin{document}

\preprint{APS/123-QED}

\title{Andreev tunneling through a double quantum-dot system coupled to a ferromagnet
and a superconductor: effects of mean field electronic correlations }

\author{E. C. Siqueira}\email{ecosta@ifi.unicamp.br}
\author{G. G. Cabrera}\email{cabrera@ifi.unicamp.br}%
\affiliation{Instituto de Física `Gleb Wataghin', UNICAMP, C.P. 6165, Campinas 13083-970, SP, Brazil}

\date{\today}
\begin{abstract}
We study the transport properties of a hybrid nanostructure composed of a ferromagnet, two
quantum dots, and a superconductor connected in series. By using the non-equilibrium Green's function
approach, we have calculated the electric current, the differential conductance and the transmittance for
energies within the superconductor gap. In this regime, the mechanism of charge transmission is the Andreev
reflection, which allows for a control of the current through the ferromagnet polarization. We have also
included interdot and intradot interactions, and have analyzed their influence through a mean field
approximation. In the presence of interactions, Coulomb blockade tend to localized the electrons at the double-dot system, leading to an asymmetric pattern for the density of states at the dots, and thus reducing the transmission
probability through the device. In particular, for non-zero polarization,
the intradot interaction splits the spin degeneracy,
reducing the maximum value of the current due to different spin-up and spin-down densities of states.
Negative differential conductance (NDC) appears for some regions of the voltage bias, as a result of
the interplay of the Andreev scattering with electronic correlations. By applying a gate voltage at
the dots, one can tune the effect, changing the voltage region where this novel phenomenon appears. This
mechanism to control the current may be of importance in technological applications.

\end{abstract}

\maketitle

\section{Introduction}

The interest in transport properties of mesoscopic systems has increased a lot due to their potential for present and future technologies. Recent advances in the experimental development of
nanostructures are mainly aimed at the study of purely quantum phenomena and effects based on electron-spin properties (\emph{spintronics}). In particular, hybrid resonant structures composed by one or more quantum dots ($QD$) coupled
to normal ($N$), ferromagnetic ($F$) and superconductor ($S$) metals have been studied \cite{reviewdoubleqdots,reviewspintronics,reviewpoucoseletrons,DasSarma,shangguan,chenpsin,sunwanglin1,japfengwujiang}. In systems composed by one
quantum dot, electron-spin properties have been extensively explored. In the special case of junctions composed by a ferromagnet and a superconductor it is possible to construct spin valves which control the current flow through those systems. Andreev reflection permits such control, by varying the polarization of the ferromagnet attached to the system, as shown in several papers
\cite{claro,Song,deutscher,Shi,fu,japTquantumdot,principal}. Andreev reflection \cite{andreevref} ($AR$) is a mechanism in which a Cooper pair is formed in the superconductor from the combination
of an incident electron coming from the normal metal with energy $\omega$ and spin $\sigma$, with another electron with energy $-\omega$ and spin $-{\sigma}$. Both electrons enter the superconductor as a Cooper pair, leaving a reflecting hole in the ferromagnetic electrode. Andreev states are located within the superconductor gap, where no quasi-particles states are available.

\begin{figure}[h]
\includegraphics[scale=0.50]{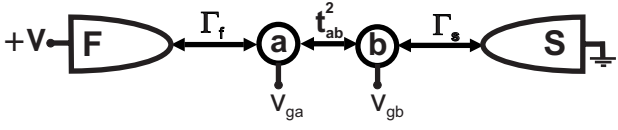}
\caption{\label{esquema}(Color Online) Schematic diagram showing the system
studied in this work. The dot coupled to the ferromagnet electrode
($F$) is called $a$, and $b$ is the one coupled
to the superconductor ($S$). The superconductor has its chemical potential fixed to
zero, and the voltage bias is applied to the ferromagnet. Gate voltages are also applied at the dots.
The different couplings are also indicated in the figure.}
\end{figure}

In this work we have studied the transport properties of a hybrid nanostructure composed by a ferromagnet, two quantum dots \cite{Hofstetter,LiJMat,japTquantumdot,pan,bergeret,WooLee,hornberger}, and a superconductor connected in series ($F-QD_{a}-QD_{b}-S$). The addition of an extra quantum dot will allow us to study the interplay of electron correlations  at the dots (for both, intra and interdot interactions), with the Andreev current. Figure \ref{esquema} shows a schematic diagram of the system. The superconductor chemical potential is fixed to zero ($\mu_{S}=0$) and the bias is applied to the ferromagnetic electrode. There are also
applied gate voltages at the dots $a$ and $b$, namely $V_{ga}$ and $V_{gb}$, respectively. By using the non-equilibrium Green's function \cite{keldysh,Jauho1,rammer,claro}, we have calculated the current ($I$),
differential conductance ($dI/dV$), Andreev transmittance ($T_{AR}$) and the local density of states (LDOS) at the dots. All quantities are calculated for energies within the superconducting gap, the relevant range for the Andreev reflection, as
functions of the voltage bias. We have also included intradot and interdot Coulomb correlations at the dots, and have analyzed its influence
on the electric current through a mean field approximation. In solids, both correlations compete to form charge or spin modulated structures.
Those symmetry broken states are not possible in finite systems, as it is the case in our double dot sample\cite{falicov}. However, dot $a$ is coupled to a ferromagnet, which breaks spin symmetry, and dot $b$ is coupled to a superconductor, which acts as a charge reservoir. Thus, interesting effects are expected, when the electronic interactions at the dots are taken into account. In this paper, those effects are displayed by the differential conductance, which shows asymmetric regions of negative values as a function of the applied bias, when the $I\times V$ characteristics are obtained.
Negative differential conductance (NDC) have been observed in hybrid nanostructures composed by normal metals \cite{ishibashi}, semiconductor based devices \cite{Mendez} and more recently in molecular Josephson junctions \cite{makk}. There are
also some theoretical studies on the NDC effect in those systems, using models beyond the mean field
approximation\cite{franssonprb,frassonjphys,lara,pedersen,Fujisawa,nguyen,zazunov}. In our work, electron interactions at the dots are treated within a mean field approach. This approximation, plus additional correlations introduced through couplings to the $F/S$ electrodes, gives rise to NDC effects. For Andreev currents, correlation parameters at the dots are limited by the size of the superconductor gap.

This paper is organized as follows: In section II we present the model under consideration and derive the
transport properties by using the non-equilibrium Green's functions. In section III the numerical results are
presented and discussed. Some conclusions are given in section IV.

\section{Model and Formulae}

\subsection{Hamiltonian}

The system displayed in figure \ref{esquema} is described by the following Hamiltonian:
\begin{align}\label{Hamiltoniano}
\mathcal{H}=\mathcal{H}_{F}+\mathcal{H}_{S}+\mathcal{H}_{dqd}+\mathcal{H}_{T},
\end{align}

\begin{align*}
\mathcal{H}_{F}=\sum_{k\sigma}(\epsilon_{k}-\sigma h-\mu_{F})a^{\dag}_{k\sigma}a_{k\sigma},
\end{align*}

\begin{align*}
\mathcal{H}_{S}=\sum_{p\sigma}\epsilon_{p}s_{p\sigma}^{\dag}s_{p\sigma}+\sum_{p}
[\Delta s^{\dag}_{p\uparrow}
s^{\dag}_{-p\downarrow}+\text{H.C.}],
\end{align*}
\begin{align*}
\mathcal{\hat{H}}_{dqd}=\sum_{\sigma}E_{a\sigma}\hat{n}_{a\sigma}+\sum_{\sigma}E_{b\sigma}\hat{n}_{b\sigma}
\end{align*}
where,
\begin{align}\label{nivel:a}
E_{a\sigma}=eV_{ga}+
\dfrac{\mathcal{K}}{2}\left\bra\hat{n}_{b}\right\ket+\dfrac{\mathcal{U}}{2}\left\bra\hat{n}_{a\bar{\sigma}}\right\ket\\
\label{nivel:b} E_{b\sigma}=eV_{gb}+
\dfrac{\mathcal{K}}{2}\left\bra\hat{n}_{a}\right\ket+\dfrac{\mathcal{U}}{2}\left\bra\hat{n}_{b\bar{\sigma}}\right\ket
\end{align}

\begin{multline*}
\mathcal{H}_{T}=\sum_{ k\sigma}[t_{f}a_{k\sigma}^{\dag}c_{a\sigma}+
\text{H.C.}]+\sum_{p\sigma}[t_{s}s_{p\sigma}^{\dag}c_{b\sigma}+
\text{H.C.}]\\+
\sum_{\sigma}[t_{ab}c^{\dag}_{a\sigma}c_{b\sigma}+\text{H.C.}].
\end{multline*}

$\mathcal{H}_{F}$ is the Hamiltonian of the ferromagnet $F$ described by the Stoner model. The spin bands
of $F$ are shifted by $h$, the exchange energy. The ferromagnet chemical potential is fixed by the applied
bias, i.e., $\mu_{F}=eV$. $\mathcal{H}_{S}$ is the Hamiltonian for a BCS superconductor with chemical
potential fixed to zero as the ground, $\mu_{S}=0$. $\mathcal{H}_{dqd}$ is the Hamiltonian for the quantum
dots in the mean field approximation, which permits an exact equation for the self-energy. The energies $E_{a\sigma}$ and $E_{b\sigma}$ are renormalized by the interactions
$\mathcal{K}$ (interdot) and $\mathcal{U}$ (intradot). The interactions also couple the renormalized
energy levels with the mean occupations $\left\bra\hat{n}_{a}\right\ket$ and $\left\bra\hat{n}_{b}\right\ket$.
In addition, it is included a gate voltage at the quantum dots $a$ and $b$, namely $V_{ga}$ and $V_{gb}$,
respectively. $\mathcal{H}_{T}$ is the Hamiltonian which describes all the tunneling processes:
between dot $a$ and the ferromagnet, with amplitude $t_{f}$, between dots with amplitude $t_{ab}$, and between dot
$b$ and the superconductor, with amplitude $t_{s}$.

\subsection{Green's functions}
To calculate the transport properties we have used the non-equilibrium Green's function method \cite{keldysh,Jauho1,rammer,claro}. All the
physical quantities can be cast in terms of the Green's function of the dots. By using the Nambu $4\times 4$ notation the retarded Green´s functions of the quantum dots are given by:
\begin{align}\label{graa}
\mathbf{G}^{r}_{aa}=\mathbf{G}_{aa}^{r0}+\mathbf{G}^{r}_{aa}\mathbf{t}^{\dag}_{ab}\mathbf{G}^{r0}_{bb}\mathbf{t}_{ab}\mathbf{G}_{aa}^{r0}
\end{align}
\begin{align}\label{grbb}
\mathbf{G}^{r}_{bb}=\mathbf{G}_{bb}^{r0}+\mathbf{G}^{r}_{bb}\mathbf{t}^{\dag}_{ab}\mathbf{G}^{r0}_{aa}\mathbf{t}_{ab}\mathbf{G}_{aa}^{r0}.
\end{align}
with,
\begin{align}\label{gaar0}
\mathbf{G}_{aa}^{r0}=\mathbf{g}^{r}_{aa}(\mathbf{1}-\mathbf{\Sigma}^{r}_{F}\mathbf{g}^{r}_{aa})^{-1}
\end{align}
\begin{align}\label{gbbr0}
\mathbf{G}_{bb}^{r0}=\mathbf{g}^{r}_{bb}(\mathbf{1}-\mathbf{\Sigma}^{r}_{S}\mathbf{g}^{r}_{bb})^{-1}.
\end{align}

In these equations $\mathbf{G}^{r}_{aa}$ is the Green's function of the quantum dot $a$; $\mathbf{G}^{r}_{bb}$ is the Green's function of the quantum dot $b$; $\mathbf{g}^{r}_{aa}$ and $\mathbf{g}^{r}_{bb}$ are the Green's functions of
the dots $a$ and $b$ isolated from the electrodes; $\mathbf{t}_{ab}$ describes the coupling between the dots; $\mathbf{\Sigma}^{r}_{F}$ and $\mathbf{\Sigma}^{r}_{S}$ are the retarded self-energies which describe the coupling of the dots with the superconductor and ferromagnet electrodes, respectively. Explicitly these self-energies are written as,
\begin{align}\label{teste}
\mathbf{\Sigma}^{r,a}_{F}(\omega)=\mp \frac{i}{2}
\begin{bmatrix}
\Gamma_{f\uparrow}    &   0   &   0   &   0   \\
0    &   \Gamma_{f\downarrow}   &   0   &   0   \\
0    &   0   & \Gamma_{f\downarrow}   &   0   \\
0    &   0   &   0   &  \Gamma_{f\uparrow}   \\
\end{bmatrix},
\end{align}
with $\Gamma_{f\sigma}=2\pi|t_{f}|^{2}N_{\sigma}$ is the coupling strength, with $t_{f}$ being the tunneling
amplitude and $N_{\sigma}$ the density of states for the ferromagnet spin $\sigma$ band; and
\begin{align}\label{selfenergy:superconductor}
\mathbf{\Sigma}^{r,a}_{S}(\omega)=\mp\frac{i}{2}\Gamma_{s}\rho(\omega)
\begin{bmatrix}
 1   &    -\dfrac{\Delta}{\omega}   &   0   &   0 \\
-\dfrac{\Delta}{\omega}   &     1  &   0   &   0 \\
0  &    0   &   1  &   \dfrac{\Delta}{\omega} \\
0  &    0   &   \dfrac{\Delta}{\omega}  & 1
\end{bmatrix},
\end{align}
where $\Gamma_{s}=2\pi |t_{s}|^{2}N_{s}(0)$, with $N_{s}$ being the density of states of the superconductor
in the normal state and $\rho_{s}$ is the modified BCS density of states  $ \rho(\omega)\equiv\dfrac{|\omega|\theta(|\omega|-\Delta)}{\sqrt{\omega^{2}-\Delta^{2}}}
+\dfrac{\omega\theta(\Delta-|\omega|)}{i\sqrt{\Delta^{2}-\omega^{2}}}$, with the imaginary part accounting for Andreev states within the gap\cite{claro,tinkhamprb}.

Besides the retarded and advanced Green's functions, it is necessary to obtain the Keldysh Green's functions,
which are calculated by the equation of motion technique. Since it is used a mean field approximation for the
interaction, the result for this Green's function is exact. The equation obtained for the Keldysh Green's function of
dot $a$ is given by:
\begin{align}\label{keldysh:recuperada}
\mathbf{G}^{<}_{aa}(\omega)&=
\mathbf{G}^{r}_{aa}(\omega)\mathbf{\Sigma}_{T}^{<}(\omega)\mathbf{G}^{a}_{aa}(\omega),
\end{align}
with the ``lesser" self-energy $\mathbf{\Sigma}_{T}^{<}$:
\begin{align}
\label{selfenergy:global}
\mathbf{\Sigma}_{T}^{<}(\omega)&=\mathbf{\Sigma}_{F}^{<}(\omega)+\mathbf{t}^{\dag}_{ab}\mathbf{G}^{r0}_{bb}\mathbf{\Sigma}_{S}^{<}(\omega)\mathbf{G}_{bb}^{a0}(\omega)
\mathbf{t}_{ab}.
\end{align}

Correspondingly, the Keldysh equation for quantum dot $b$ is given by:
\begin{align}\label{keldyshb:recuperada}
\mathbf{G}^{<}_{bb}(\omega)&=\mathbf{G}^{r}_{bb}(\omega)\mathbf{\Sigma}^{<}_{Tb}(\omega)\mathbf{G}_{bb}^{a}(\omega),
\end{align}
with the ``lesser" self-energy $\mathbf{\Sigma}_{Tb}^{<}$:
\begin{align}\label{selfenergydotb:global}
\mathbf{\Sigma}^{<}_{Tb}(\omega)&=\mathbf{\Sigma}_{S}^{<}(\omega)+
\mathbf{t}_{ab}\mathbf{G}^{r0}_{aa}\mathbf{\Sigma}^{<}_{F}(\omega)\mathbf{G}_{aa}^{a0}(\omega)\mathbf{t}^{\dag}_{ab}.
\end{align}

The ``lesser" self-energy for the ferromagnet electrode is given by:
\begin{align}\label{sigma1}
\mathbf{\Sigma}_{F}^{<}(\omega)=i
\begin{bmatrix}
f_{F}\Gamma_{f\uparrow} &  0 & 0 &  0 \\
0 &  \bar{f_{F}}\Gamma_{f\downarrow} & 0 &  0 \\
0 &  0 & f_{F}\Gamma_{f\downarrow} &  0 \\
0 &  0 & 0 &  \bar{f_{F}}\Gamma_{f\uparrow} \\
\end{bmatrix},
\end{align}
in which $f_{F}=f(\omega-eV)$ and $\bar{f_{F}}=(\omega+eV)$ are the Fermi functions for electrons and holes,
respectively.

The ``lesser" self-energy for the superconductor electrode is given by:
\begin{align}\label{sigmas}
\boldsymbol{\Sigma}_{S}^{<}(\omega)=if\Gamma_{s}\tilde{\rho}(\omega)
\begin{bmatrix}
 1   &    -\dfrac{\Delta}{\omega}   &   0   &   0 \\
-\dfrac{\Delta}{\omega}   &     1  &   0   &   0 \\
0  &    0   &   1  &   \dfrac{\Delta}{\omega} \\
0  &    0   &   \dfrac{\Delta}{\omega}  & 1
\end{bmatrix},
\end{align}
where $f_{S}=f(\omega)$ is the Fermi function for the superconductor electrode and
$\tilde{\rho}=\dfrac{|\omega|}{\sqrt{\omega^{2}-\Delta^{2}}}$ is the conventional BCS density of states.

Equation \eqref{selfenergy:global} shows that the dot $a$, which is coupled to the ferromagnetic electrode
on its left side, `sees' on its right side an effective electrode as a result of the interplay of dot $b$ with the
superconductor. Equation \eqref{selfenergydotb:global} can be interpreted in similar terms for dot $b$,
with a `bare' superconductor electrode on  the right side, and an effective electrode on the left,
resulting from the interaction of dot $a$ with the ferromagnet.
Since the superconductor and the ferromagnet present different band structures, there is an
intrinsic asymmetry in this system which manifests itself in the transport properties.

\subsection{Physical Quantities}

The Green's functions of the last section, calculated by the equation of motion method, permit to determine all
the physical quantities necessary to analyze the transport properties of the $F-QD_{a}-QD_{b}-S$ system.
Since the interaction couples the dot levels through the mean occupation, as shown by equations
\eqref{nivel:a} and \eqref{nivel:b}, it is necessary to perform a self-consistent calculation to determine
the occupation at the dots first. Then, one can proceed to calculate the physical quantities of interest.

In the following we show the expressions we have used to compute the LDOS, the current, the transmittance and the mean
occupation.

\subsubsection{Local density of states (LDOS)}

The LDOS of the quantum dots comes from the matrix elements [11] and [33] of the retarded Green's function
matrix (electron components in Nambu space). The LDOS for dots $a$ and $b$  are, respectively:
\begin{align}\label{ldos:A}
\text{LDOS-A}&= -\dfrac{1}{\pi}\text{Im}[G^{r}_{aa,11}+G^{r}_{aa,33}]\\\label{ldos:b} \text{LDOS-B} &=
-\dfrac{1}{\pi}\text{Im}[G^{r}_{bb,11}+G^{r}_{bb,33}].
\end{align}

\subsubsection{Transmittance and current}

Since the current is conserved, it can be calculated at any point of the circuit. Here, we choose to calculate the current at the ferromagnetic electrode, as the temporal variation of the number of electrons , i.e.:
\[I=-e\left\bra\dfrac{d\hat{N}_{F}}{dt}\right\ket\ ,
\]
where $\hat{N}_{F}=\sum_{k\sigma}a^{\dag}_{k\sigma}a_{k\sigma}$. By using the Heisenberg equation and the
definition of the ``lesser" Green's function of the dot $a$, it's possible to write the current as follow:
\begin{align}\label{corrente:finalexpression}
I=\frac{e}{\hbar}\int d\omega\left[
\mathbf{G}^{r}_{aa}(\omega)\mathbf{\Sigma}_{F}^{<}(\omega)+\mathbf{G}^{<}_{aa}(\omega)\mathbf{\Sigma}_{F}^{a}(\omega)+\text{H.C.}\right]_{11+33}\ ,
\end{align}
where the index $11+33$ indicates a sum over the electron components in the Nambu space matrix. By
substituting the matrix elements, the current can be cast to the following form:
\begin{align}\label{corrente:defintiva} I=\frac{e}{h}\int
d\omega~A(\omega)(f_{F}-\bar{f_{F}}).
\end{align}

In this work we only consider Andreev transport, for energies within the superconductor gap.
Thus, the current amplitude corresponding to the contribution of quasi-particles tunneling is zero.

The expression for the amplitude $A(\omega)$ is given by:
\begin{multline*}
A=\Gamma_{f\uparrow}\left(|G^{r}_{aa,14}|^{2}\Gamma_{f\uparrow}+|G^{r}_{aa,12}|^{2}\Gamma_{f\downarrow}\right)+\\
\Gamma_{f\downarrow}\left(|G^{r}_{aa,34}|^{2}\Gamma_{f\uparrow}+
|G^{r}_{aa,32}|^{2}\Gamma_{f\downarrow}\right).
\end{multline*}

The transmittance is obtained from the current formula:
\begin{align}\label{tar}
T_{AR}=\dfrac{1}{2}[\Gamma_{f\uparrow}\left(|G^{r}_{aa,14}|^{2}\Gamma_{f\uparrow}+|G^{r}_{aa,12}|^{2}\Gamma_{f\downarrow}\right)+\\\nonumber
\Gamma_{f\downarrow}\left(|G^{r}_{aa,34}|^{2}\Gamma_{f\uparrow}+
|G^{r}_{aa,32}|^{2}\Gamma_{f\downarrow}\right)].
\end{align}

\subsubsection{Self-consistent calculations}

Since the Green's functions are dependent on the mean occupations via equations \eqref{nivel:a} and
\eqref{nivel:b}, it is necessary to calculate those quantities at the dots. From the definition of the
``lesser" Green's function, one straightforwardly obtains the system of equations below:
\begin{align*}
\left\bra n_{a\uparrow}\right\ket=\dfrac{1}{2\pi i}\int_{-\infty}^{+\infty}G^{<}_{aa,11}(\omega)\\
\left\bra n_{a\downarrow}\right\ket=\dfrac{1}{2\pi i}\int_{-\infty}^{+\infty}G^{<}_{aa,33}(\omega)
\end{align*}
\begin{align*}
\left\bra n_{b\uparrow}\right\ket=\dfrac{1}{2\pi i}\int_{-\infty}^{+\infty}G^{<}_{bb,11}(\omega)\\
\left\bra n_{b\downarrow}\right\ket=\dfrac{1}{2\pi i}\int_{-\infty}^{+\infty}G^{<}_{bb,33}(\omega)
\end{align*}

These integral equations have to be solved numerically in a self-consistent way. Once the occupation
numbers are obtained, it is possible to calculate the other physical quantities. Results are shown below.

\section{Results and discussion}

Next, we present the results obtained from numerical calculations. Firstly, we show the local density of states
(LDOS) at the quantum dots in the absence of electronic correlations. We investigate the effects of the
different couplings of the model, namely the coupling between dots and the coupling of the dots with the electrodes.
In the following, we discuss the role of the interdot interaction. In the specific case with no polarization in the
ferromagnet, $P=0$, we have  observed the appearance of the NDC effect, for some values of
the interaction. The inclusion of the intradot interaction lifts the spin degeneracy, as can
be seen by the splitting of the differential conductance peaks. In this case the NDC also appears,
but is reduced with the increase of the polarization.

\subsection{Noninteracting case}

In this section, the LDOS of the quantum dots is described without interactions. This permits us to analyze the
resonance structure presented by these quantities, which plays a central role in all transport properties,
including the interacting case as well.
The system is asymmetric, since the ferromagnetic left electrode is modeled with a continuous density of states,
while the superconductor electrode, on the right side, presents a gap for quasi-particles states, with a complex
density of states within the gap, corresponding to evanescent Andreev states, responsible for the Copper pair
conversion at the interface.

\begin{figure*}[h]
\includegraphics[scale=0.77]{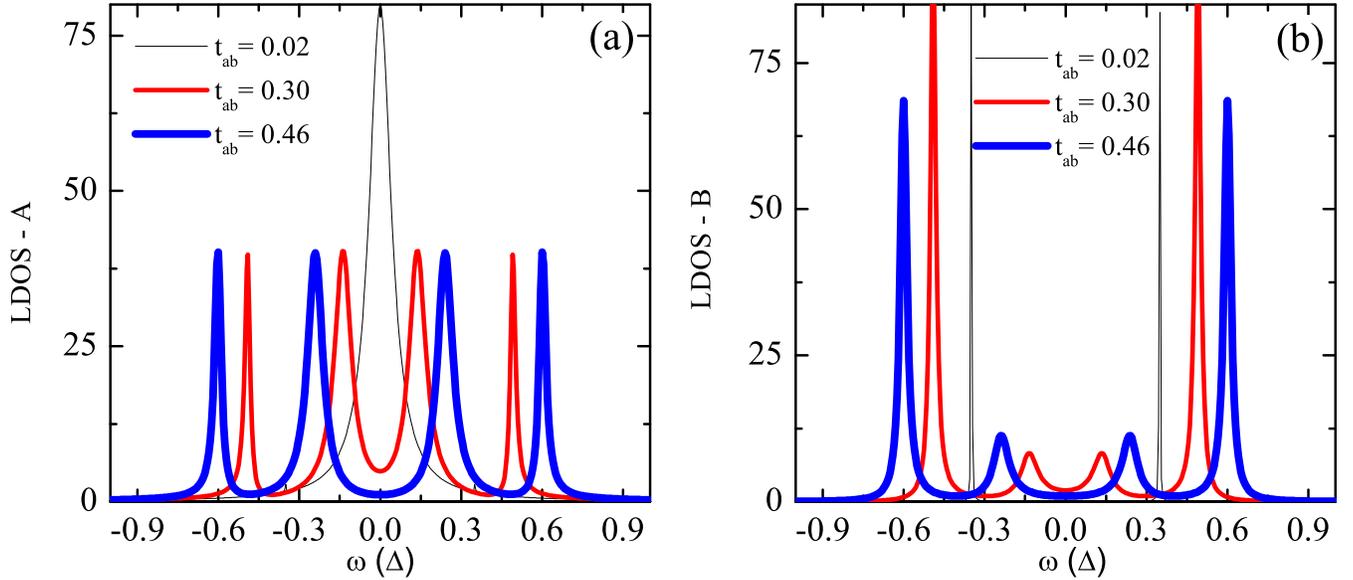}
\caption{\label{LDOSnInt:tab}(Color Online) LDOS for different values of coupling between dots, $t_{ab}$. Fixed
parameters: $\Gamma_{f}=0.1$, $\Gamma_{s}=1.0$, and $P=0$, $k_{B}T=0.01$ and $eV_{ga}=eV_{gb}=0$.
(a) LDOS for QD $a$, coupled to
the ferromagnet. For $t_{ab}=0.02$, the LDOS is dominated by the coupling with the ferromagnet, revealed by the
broadening of the central peak. By increasing $t_{ab}$, the resonances from the superconductor (external peaks)
and from interdot coupling (central peaks) appear. (b) LDOS for QD $b$, coupled to the superconductor.
For $t_{ab}=0.02$ the LDOS is dominated by the coupling with the superconductor, as displayed by the equidistant
peaks around $\omega=0$; these peaks are the Andreev resonances. By increasing $t_{ab}$, the interdot coupling
peaks (central peaks) appear in addition to the Andreev resonances. All parameters are expressed in units of the
superconductor gap.}
\end{figure*}

Figure \ref{LDOSnInt:tab} shows the LDOS at the dots, for different values of the interdot
coupling (electron hopping between dots) $t_{ab}$, in units of the superconducting gap.
For $t_{ab}=0.02$, the dots are almost decoupled from each other. As a
result, features of the density of states mainly reflect the coupling with the electrodes.
LDOS for dot $a$ presents one peak centered in $\omega=0$ with a finite width. The broadening
results from the hybridization of the dot level with the ferromagnetic band. There is a finite
probability for the electron to escape from the dot to the electrode.
On the other hand, LDOS-B presents two sharp symmetrical peaks. This resonant structure represents
the hybridization between the dot level with the Andreev states. The peaks corresponds to the electron and hole
channels, as expected from the BCS model for the superconductor electrode.

As the coupling between the dots is increased, two additional peaks at the center emerge for both LDOS,
as observed in the examples for $t_{ab}=0.30$ and $0.46$. These peaks come from the resonance between the
discrete dot levels. For LDOS-B, the intensity of the Andreev peaks decay with the dot coupling.
\begin{figure*}[h]
\includegraphics[scale=0.77]{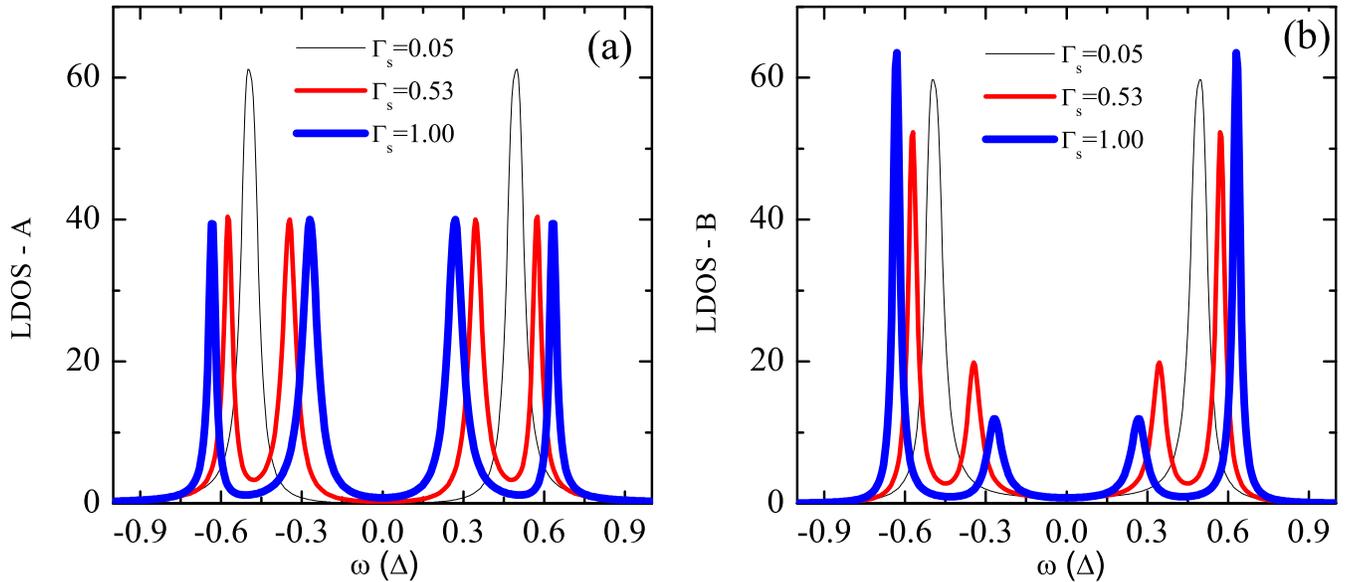}
\caption{\label{LDOSnInt:gamas}(Color Online) LDOS for different values of the coupling with the superconductor, $\Gamma_{s}$.
Fixed parameters: $\Gamma_{f}=0.1$,  $t_{ab}=0.5$, $P=0$, $k_{B}T=0.01$ and $eV_{ga}=eV_{gb}=0$.
(a) LDOS for quantum dot $a$, coupled to
the ferromagnet. (b) LDOS for quantum dot $b$, coupled to the superconductor. The distinct behavior of
LDOS-A and  LDOS-B is explained by the stronger coupling of $b$ with $S$. The increase in the coupling with $S$ results in a bigger
separation between the resonance peaks. All parameters are expressed in units of the
superconductor gap.}
\end{figure*}
\begin{figure*}[t]
\includegraphics[scale=0.8]{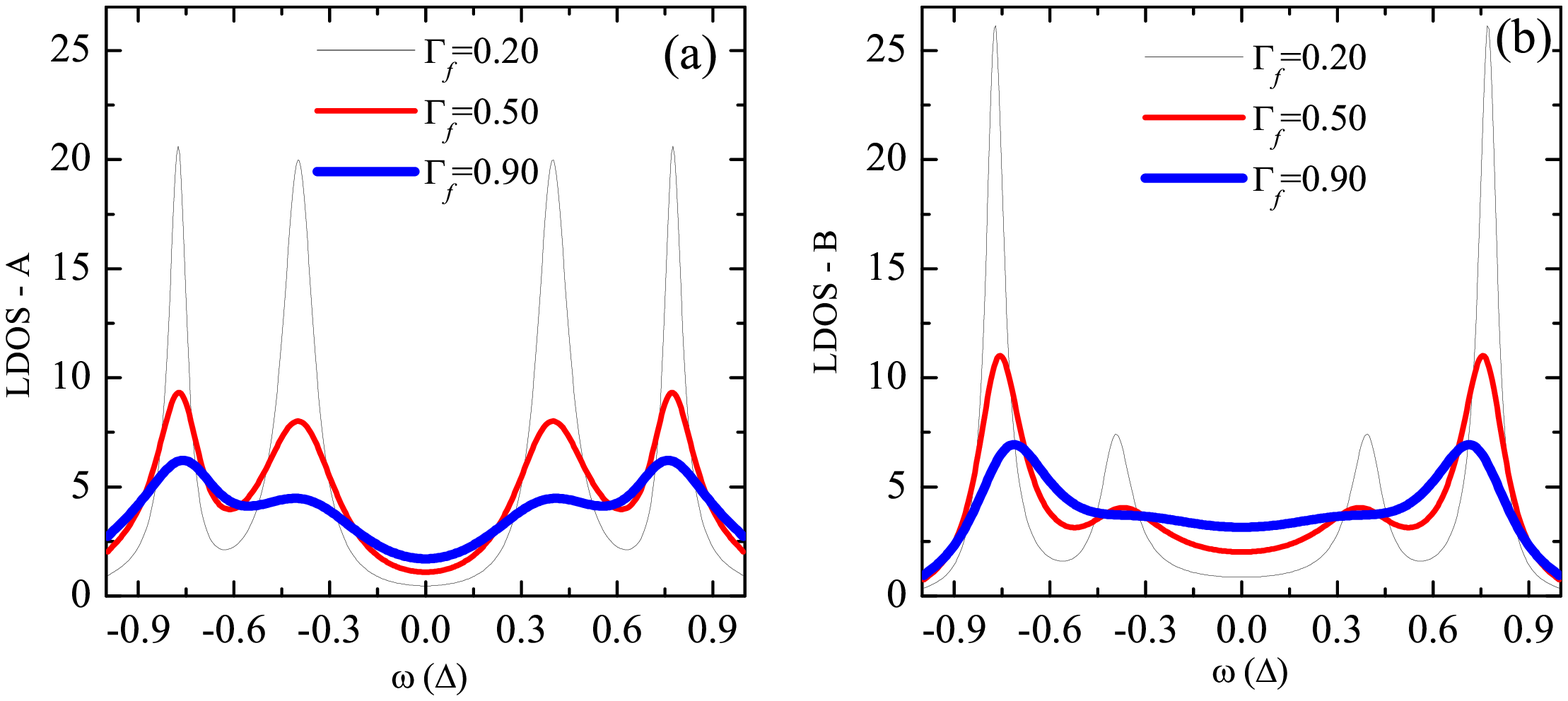}
\caption{\label{LDOSnInt:gama1}(Color Online) LDOS for different values of coupling with the ferromagnet, $\Gamma_{f}$.
Fixed parameters: $\Gamma_{s}=1.0$, $t_{ab}=0.5$, $P=0$, $k_{B}T=0.01$ and $eV_{ga}=eV_{gb}=0$.
(a) LDOS for quantum dot $a$, coupled to
the ferromagnet. (b) LDOS for quantum dot $b$, coupled to superconductor. By increasing $\Gamma_{f}$, the density
of states presents a broader pattern, displaying an admixture between the ferromagnetic energy band
and the hybridized states of the dots. All parameters are expressed in units of the superconductor gap.}
\end{figure*}

The effect of the coupling with the superconductor is illustrated
by the LDOS curves shown in figure \ref{LDOSnInt:gamas}. When
the interaction with the superconductor is weak $\Gamma_{s}=0.05$,
both LDOS present a two-peaks structure resulting from the
interdot coupling ($t_{ab}=0.5$ in the examples shown). When the coupling with the superconductor is
increased, the Andreev peaks appear and are more intense in LDOS-B.

The effect on the LDOS by varying the coupling with ferromagnetic electrode is shown in
figure \ref{LDOSnInt:gama1}. When the coupling with the ferromagnet is increased, the discrete structure of
the LDOS is transformed into a continuum of states, as a result of the hybridization of the discrete dot levels
with the continuous band of the ferromagnet. Internal peaks almost disappear for $\Gamma_{f}>0.5$.

Isolated  quantum dots present one level degenerate in spin. When coupled to each other with $t_{ab}$,
there is an admixture of them, resulting in a bonding and an anti-bonding levels, in analogy with a $H_{2}$ molecule
\cite{falicov}. In our model, those levels corresponds to the central peaks of the LDOS. When the electrodes
are attached to the double dot system, the above peaks broaden and two additional peaks appear corresponding to
the superconducting Andreev states. By tuning the parameters of the model, it is possible to change the number of peaks,
their widths and the distance between them, which in turn can be used to control the current.

\subsection{Interacting case: Inter-dot interaction}

\begin{figure*}[H]
\includegraphics[scale=0.77]{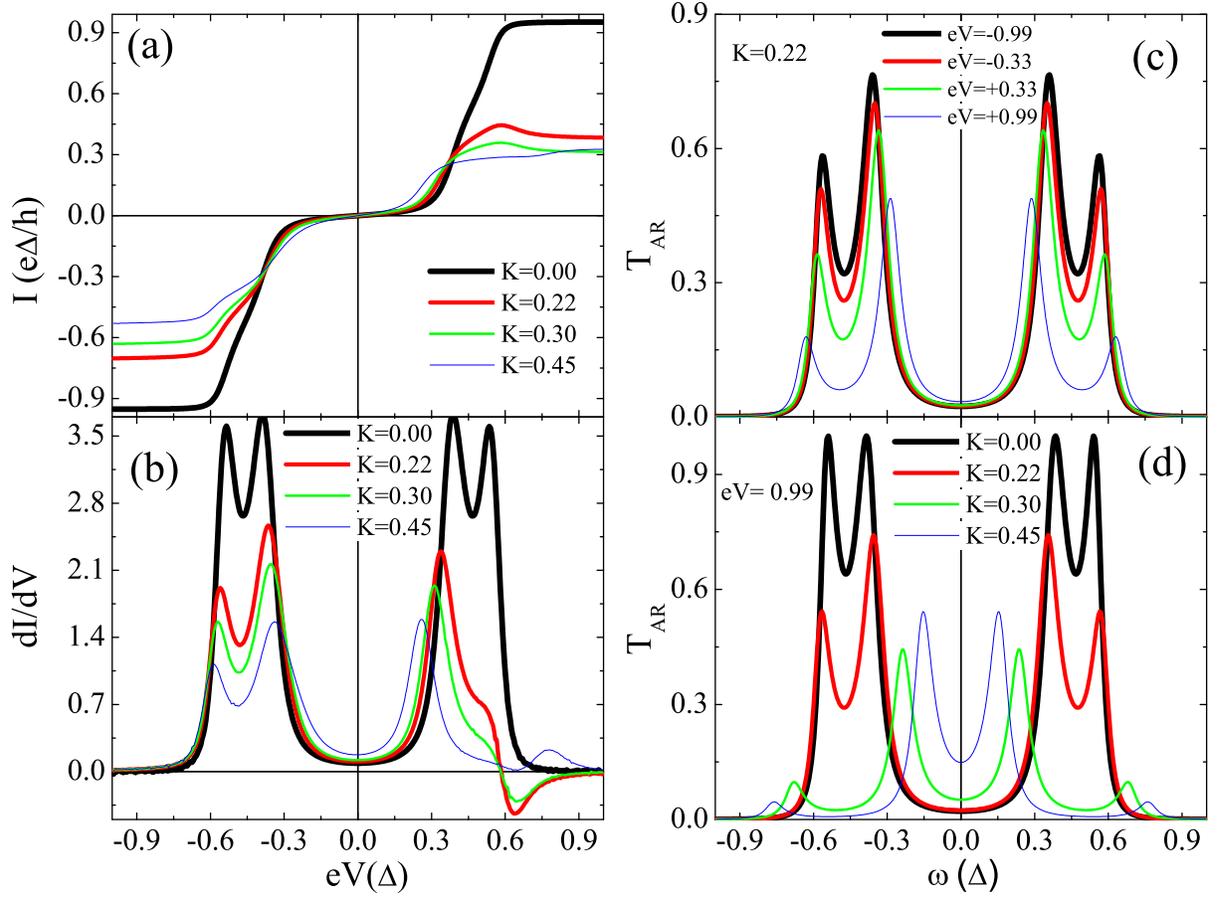}
\caption{\label{variando:K}(Color Online)  (a) current ($I$) versus applied bias ($V$) for some values of interdot
interaction. (b) Corresponding differential conductance, showing regions of negative
values. (c) Andreev transmittance ($T_{AR}$) for some values of applied bias for $\mathcal{K}=0.22$. (d)
Andreev transmittance ($T_{AR}$) for some values of $\mathcal{K}$, for applied bias $eV=0.99$. Fixed
parameters: $\Gamma_{f}=0.19$, $\Gamma_{s}=0.40$, $t_{ab}=0.5$, $P=0$, $\mathcal{U}=0$, $k_{B}T=0.01$ and
$eV_{ga}=eV_{gb}=0$. All parameters are expressed in units of the superconductor gap.}
\end{figure*}

In figure \ref{variando:K}(a) we plot some $I\times V$ characteristics, for different values of the interdot
interaction $\mathcal{K}$. These curves show a plateau pattern which is  due to the peak structure of the LDOS
at the quantum dots. When the interaction is increased, the plateau value is
reduced, ranging from $I=0.90$ for $\mathcal{K}=0$ to $I=0.30$ for $\mathcal{K}=0.45$,
since higher values of the interaction implies a stronger Coulomb repulsion between
dots. But for small voltages ($eV<0.30$), we observe an unusual behavior, where the trend is inverted,
although this is a tiny effect.
In figure \ref{variando:K}(b), we plot the corresponding differential conductance, which
allow for a better resolution of the $I\times V$ curves. The symmetric structure for $\mathcal{K}=0$
is broken when $\mathcal{K}\neq 0$, the asymmetry being more pronounced the higher the values of $\mathcal{K}$.
For some examples of the figure, NDC in the characteristics is found around $\mathcal{K}= 0.6$.
From our numerical calculations, NDC effects are present in the range $0.08<\mathcal{K}<0.4$.
For $\mathcal{K}$ greater than $0.4$, NDC is suppressed and a positive peak emerges in $dI/dV$,
as can be seen in the example for $\mathcal{K}=0.45$.

From these results, we conclude that the mechanism of the NDC is not linearly related to the Coulomb blockade effect.
The interaction plays a more subtle role in changing the transmittance of the system. In fact,
looking at the differential conductance,
we note that when increasing $\mathcal{K}$, the second peak for positive bias is
suppressed. Thus, for some values of $\mathcal{K}$ there is a suppression of some of the
resonant peaks, and this causes an additional reduction of the transmittance for some values of the applied bias.
This effect causes the differential conductance to assume negative values.
\begin{figure*}
\includegraphics[scale=.77]{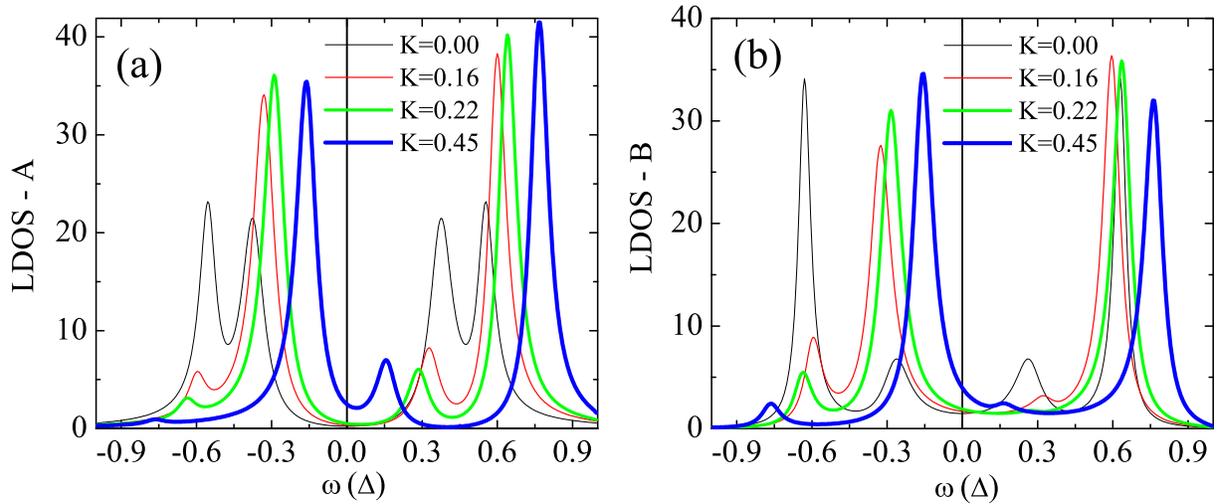}
\caption{\label{LDOS:variando K}(Color Online) LDOS for different values of interdot interaction. Fixed parameters:
$\Gamma_{f}=0.19$, $\Gamma_{s}=0.40$, $t_{ab}=0.5$,  $P=0$, $\mathcal{U}=0$, $k_{B}T=0.01$, $eV=+0.62$
and $eV_{ga}=eV_{gb}=0$. (a) LDOS for quantum dot $a$, coupled to ferromagnet. (b) LDOS for quantum dot $b$,
coupled to superconductor. In both, the interaction introduces an asymmetry which is related to the NDC effect.
All parameters are expressed in units of the superconductor gap.}
\end{figure*}
The Andreev transmittance is displayed in figures \ref{variando:K}(c) and (d). There is a variation with the applied
bias, in contrast to the non-interacting case.
The interaction couples the occupation number at the dots, which implies a non trivial dependance of the
transmittance with the applied bias. In figure \ref{variando:K}(c), we plot the transmittance for
$\mathcal{K}=0.22$, for some values of the applied bias. There is a reduction of the amplitude with the
increase of the bias, but the spectrum is symmetric with respect to $\omega$. In figure \ref{variando:K}(d),
we show the transmittance at fixed bias $eV=0.99$, for various values of $\mathcal{K}$.
There is a reduction of the transmittance and a shift of the peaks, however the variation is not systematic,
as shown by the example for $\mathcal{K}=0.45$, which does not follow the trend of the other values.
Reduction of the transmittance with increasing bias is one of the causes of NDC.
However, the absence of NDC for negative bias strongly hints that there are additional ingredients
to explain the phenomenon. One important factor is the asymmetry
in the LDOS, which appears when the interaction $\mathcal{K}$ is turned on. In figures \ref{LDOS:variando
K}(a) and (b), we show the effect of the interaction on the peak structure of the LDOS at the dots.
The symmetric shape seen in figures \ref{LDOSnInt:tab}, \ref{LDOSnInt:gamas}, and \ref{LDOSnInt:gama1} is lost
when the interaction is included.
By increasing the interaction towards the gap value, some
peaks are suppressed (one central and one external) and other are reinforced (one central and one external),
the LDOS presenting a more localized character. Central peaks of the LDOS are associated to states resonating
between dots, while external peaks are the channels for the Andreev reflection. The symmetry is critical to
allow electronic transfer through the structure, since the sum of the energies of the electrons available to form
a Cooper pair have to be
equal to the chemical potential of the superconductor (which is zero). Thus, the Andreev current is optimized when
the LDOS peaks are symmetric, and the suppression of one of them causes an effective reduction of the
current, with the emergence of the NDC effect.

When a negative bias is applied to the ferromagnetic electrode, its chemical potential is reduced in relation to the
superconductor's one. Thus, the current is established by extracting Cooper pairs from
superconductor electrode. Those electrons, with antiparallel spins, fill the lower energy states available at the dots,
so only the peaks of the negative frequency branch of the LDOS will participate in the conduction process.
This is the explanation for the absence of NDC for negative bias in the $I\times V$ characteristics, as shown in figure \ref{variando:K}(a).

\begin{figure}[h]
\includegraphics[scale=0.85]{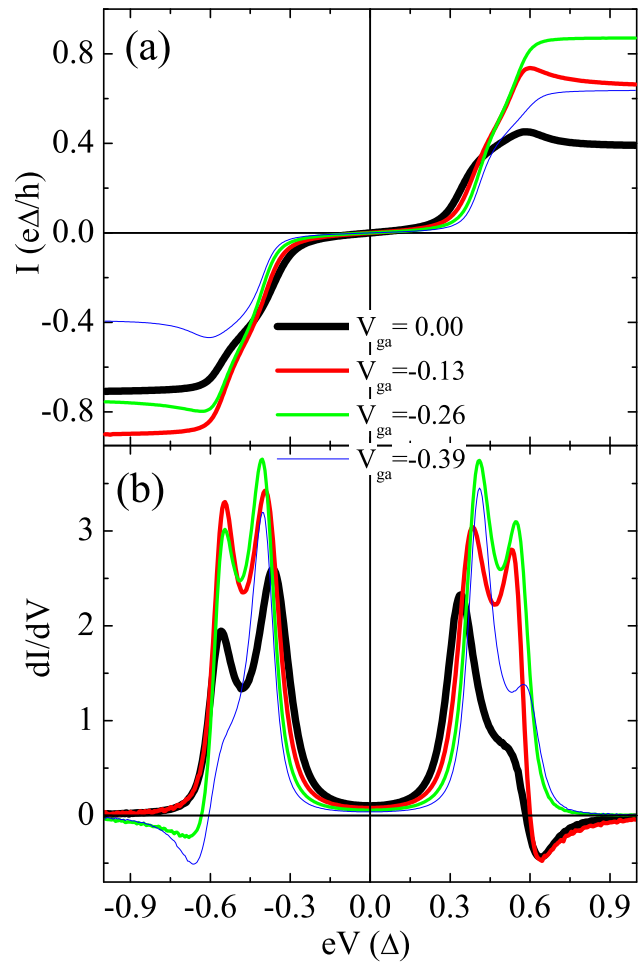}
\caption{\label{IxVVga}(Color Online) Current and differential conductance vs. applied bias for some values of the gate voltage at dot $a$. Fixed parameters: $\Gamma_{f}=0.2$, $\Gamma_{s}=0.4$, $t_{ab}=0.50$, $P=0$, $\mathcal{K}=0.22$,
$\mathcal{U}=0$, $k_{B}T=0.01$ and $V_{gb}=0$. (a) current versus applied bias: the gate potential modifies
the current profile appearing some regions of NDC in both, negative and positive values of the applied bias.
(b) Differential conductance for the corresponding $I\times V$ curves. All parameters are
expressed in units of the superconductor gap.}
\end{figure}
\begin{figure}[h]
\includegraphics[scale=.85]{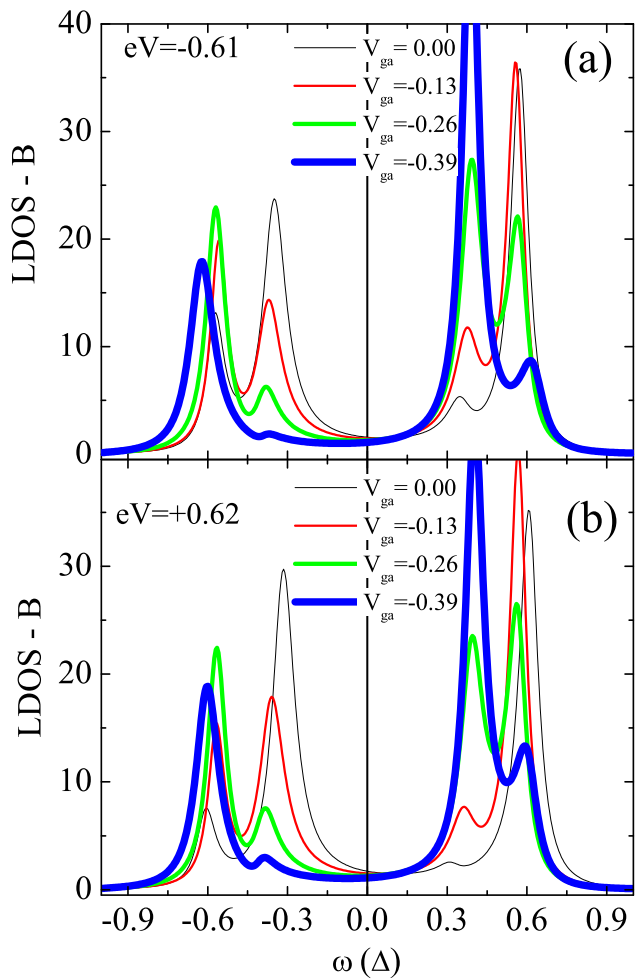}
\caption{\label{LDOS: vgatea}(Color Online) LDOS of dot $b$ vs. electron energy for some values of the gate voltage at dot $a$. Fixed parameters: $\Gamma_{f}=0.2$, $\Gamma_{s}=0.4$, $t_{ab}=0.50$, $P=0$, $k_{B}T=0.01$,
$V_{gb}=0$, $\mathcal{K}=0.22$ and $\mathcal{U}=0$. (a) LDOS-B for applied bias $eV=-0.61$.
The second peak for negative energy is progressively suppressed and vanishes at $eV_{ga}=-0.39$.
(b) LDOS-B for applied bias $eV=+0.62$. The first peak for positive energy, that was absent for $eV_{ga}=0$,
emerges with application of the gate voltage. All parameters
are expressed in units of the superconductor gap.}
\end{figure}

Next, to show that the NDC effect originates from the asymmetry of the LDOS, we have recalculated the
$\mathcal{K}=0.22$ case of figure \ref{variando:K}, but now applying a gate voltage at dot $a$, while
keeping the gate voltage at the other dot fixed and equal to $\mu_{S}=0$. The results are qualitatively similar
if the gate potential at dot $b$ is varied, while the one at $a$ is kept fixed and equal to $\mu_{S}$.
As shown in figures \ref{IxVVga}(a) and (b), the NDC appears for negative values of the applied bias,
for $V_{ga}$ approximately ranging from $-0.26$ to $-0.39$.
In the range $-0.13<V_{ga}<+0.13$, the NDC appears for positive bias.
To make contact with the asymmetry of the LDOS at the dots, in figures
\ref{LDOS: vgatea}(a) and (b), we plot the LDOS-B for values of the bias at the threshold of the
NDC,  namely $eV=-0.61$ and $+0.62$. The LDOS-A presents a similar behavior.
By tuning the gate voltage $V_{ga}$, we can change the amplitude and position of the peaks in the LDOS.
Eventually, some of the peaks vanish, given rise to the NDC effect. The above figures corroborate the role of the LDOS in the
appearance of the NDC regions. In fact, peaks of the LDOS and transmittance are resonances resulting from the coupling
between the dots and the electrodes. As shown in figures \ref{variando:K} and \ref{LDOS:variando K}, the interdot interaction
affects the LDOS and the transmittance in a way similar to ``destructive interference", changing the position and amplitude
of the peaks. One can tune such ``destructive interference", by introducing a gate voltage, thus modifying the values of bias where NDC takes place, as shown in figure \ref{LDOS: vgatea}. This process of controlling the current through the device, may be important for practical applications.

\begin{figure}[h]
\includegraphics[scale=0.85]{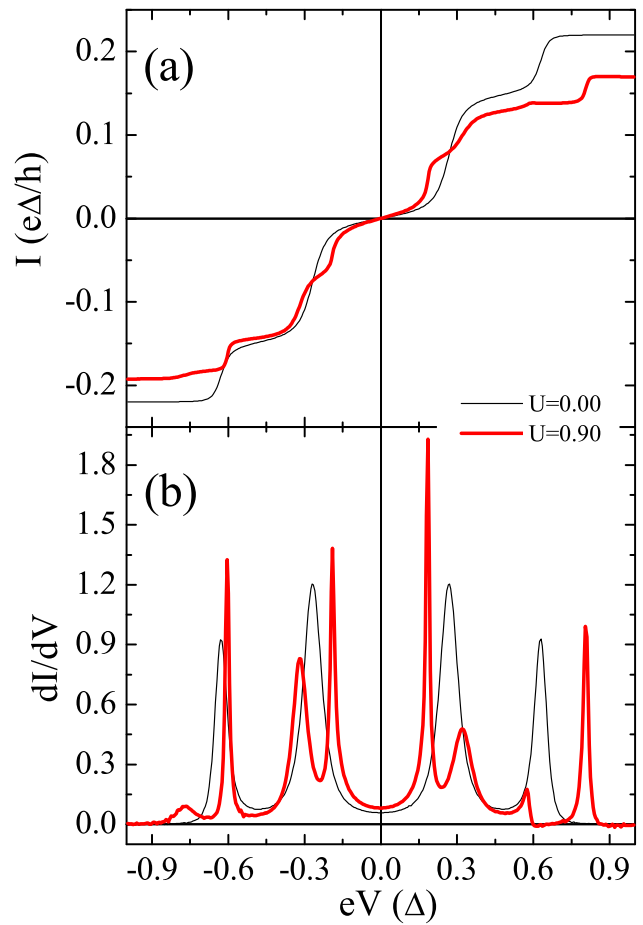}
\caption{\label{Uspliting}(Color Online) Current and differential conductance vs. applied bias for some values of the
intradot interaction. Fixed parameters: $\Gamma_{f}=0.1$, $\Gamma_{s}=1.00$, $t_{ab}=0.50$, $P=0.80$,
$\mathcal{K}=0$, $k_{B}T=0.01$ and $eV_{ga}=eV_{gb}=0$. (a) Current vs. applied bias. (b) Differential
conductance. The intradot interaction lifts the spin degeneracy, producing a splitting of the
peaks in the $dI/dV$ curves. Very different values of $\Gamma_{f}$ and $\Gamma_{s}$ have been used
in the example, to get a larger separation between the resonance peaks. All parameters are expressed in
units of the superconductor gap.}
\end{figure}

\subsection{Interacting case: Intradot interaction}

Finally, we present mean field results that include the intradot (onsite) interaction $\mathcal{U}$,
with no interdot repulsion ($\mathcal{K}=0$). As shown in equations
\eqref{nivel:a} and \eqref{nivel:b}, the intradot interaction splits the up and down-spin states at each
quantum dot, with the corresponding splitting of the transmittance and differential conductance peaks.

\begin{figure}[b]\centering
\includegraphics[scale=.85]{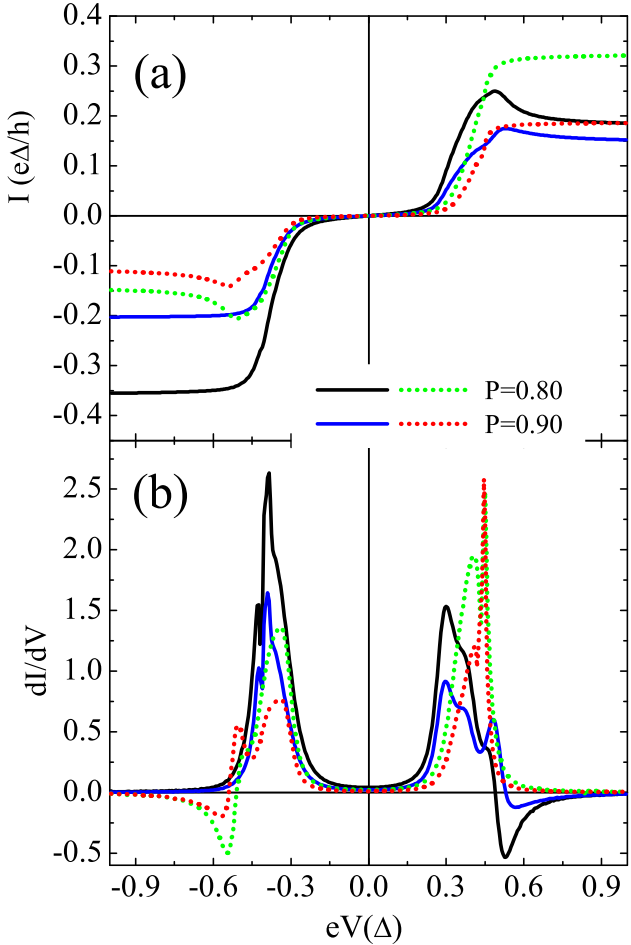}
\caption{\label{varpolarizacao}(Color Online) Current and differential conductance vs. applied bias for different values of the ferromagnet polarization. Solid lines: $V_{ga}=V_{gb}=-0.13$. Dotted lines: $V_{ga}=-0.39$~and~$ V_{gb}=-0.10$. Fixed parameters: $\Gamma_{f}=0.20$, $\Gamma_{s}=0.26$, $t_{ab}=0.40$,
$\mathcal{K}=0.25$, $\mathcal{U}=0.25$ and $k_{B}T=0.01$. (a) Current vs. applied bias. (b)
Differential conductance. The increase of the polarization suppresses NDC by reducing the
available states in the conduction process. All parameters are expressed in units of the superconductor gap.}
\end{figure}
\begin{figure}[b]\centering
\includegraphics[scale=.88]{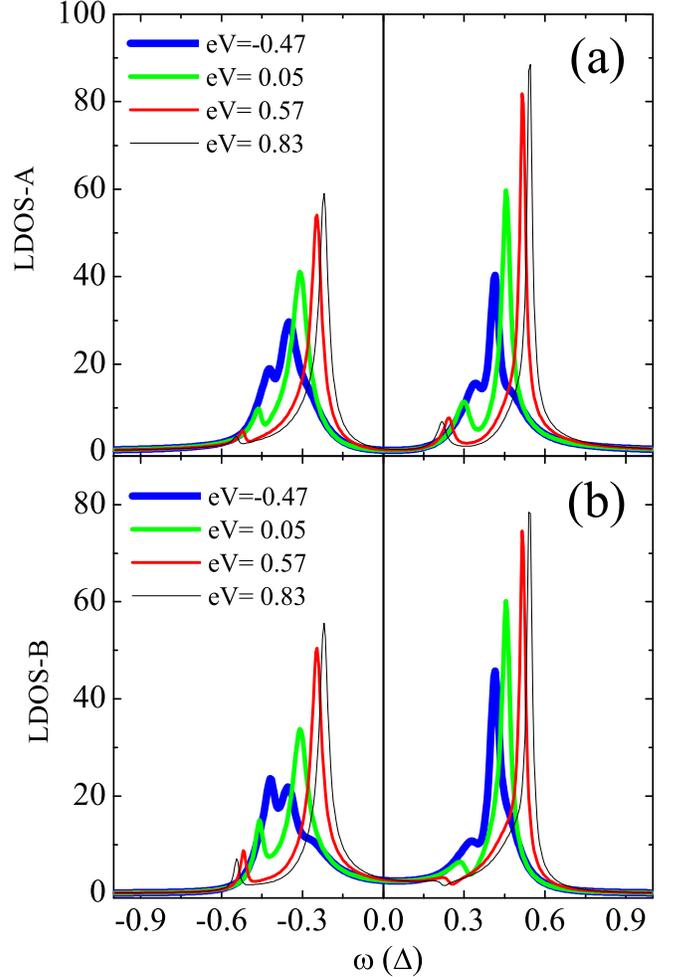}
\caption{\label{LDOSP080}(Color Online) Corresponding LDOS for some values of the applied bias. By adjusting the hopping $t_{ab}$ and the coupling with superconductor $\Gamma_{s}$, it is possible to reduce the peaks of the LDOS allowing to observe the NDC even at high values of the polarization. Fixed parameters: $P=0.80$, $\Gamma_{f}=0.20$, $\Gamma_{s}=0.26$, $t_{ab}=0.40$,
$\mathcal{K}=0.25$, $\mathcal{U}=0.25$, $k_{B}T=0.01$ and $V_{ga}=V_{gb}=-0.13$. (a) LDOS for dot $a$, coupled to ferromagnet. (b) LDOS for dot $b$, coupled to superconductor. All parameters are expressed in units of the superconductor gap.}
\end{figure}
However, as indicated by equations \eqref{nivel:a} and \eqref{nivel:b}, this effect can only be observed
for different up and down-spin occupations. This condition is met for non-zero values of the polarization $P$
of the ferromagnet, when different numbers of spin up and spin down electrons are injected to the dots.
The rates at which electrons are injected are $\Gamma_{f\uparrow}/h=\Gamma_{f}(1+P)/h$ and
$\Gamma_{f\downarrow}/h=\Gamma_{f}(1-P)/h$, for spin up and down respectively.
In figure \ref{Uspliting} the $I\times V$ characteristics
and the corresponding differential conductance are shown for different values of the intradot
interaction. As long as $\mathcal{U}\neq 0$, the
peaks start to split, and for $\mathcal{U}=0.90$ the differential conductance presents a clear pattern with eight
peaks. The $I\times V$ characteristic, for $\mathcal{U}=0.90$, also shows a number of additional steps and a final
plateau with a reduced value of the current.
The reduction of the maximum value of the current with $P$ is explained by the reduction of the available conducting channels, as discussed in reference \onlinecite{beenakker}. Since the current is established by Andreev reflection, it is necessary an equal number of spin-up and spin-down electrons to form Cooper pairs. Since the density of states for spin down is smaller, the current is limited by the number of spin-down electrons.

In the examples presented in figure \ref{Uspliting}, which corresponds to $P=0.90$,
NDC effects are absent. By increasing the polarization the NDC region is reduced, and eventually disappears, when we further increase the polarization. The mechanism that accounts for the NDC effect for intradot interaction is the same as the one presented in the previous sections: the reduction of the transmittance with the applied bias, combined with asymmetries of the LDOS. For values of the polarization of $\approx 0.90$, the mean number of electrons participating in the conduction is so reduced, that a further reduction of the channels does not imply in a reduction of the electrical current. This is the cause of the absence of NDC in the examples of the figure \ref{Uspliting}.

\subsection{Interaction case:\\ Intradot and Interdot interaction}

In the last sections we have presented NDC results with one of the interactions absent. However, it is possible to observe NDC when both interactions are active. In the figure \ref{varpolarizacao} is shown some $I\times V$ curves for $\mathcal{U}=\mathcal{K}=0.25$. In spite of the high polarization values, $P=0.80$ and 0.90, there are regions of NDC in all these curves. For $eV_{ga}=eV_{gb}=-0.13$ (solid curves), NDC appears for applied bias $eV\gtrsim 0.48$. For $V_{ga}=-0.39$~and~$ V_{gb}=-0.10$ the NDC appears for negative bias $eV\lesssim-0.48$ (dotted curves).  Besides the interactions and gate voltage values, another difference from the figure \ref{Uspliting} are the values of the coupling constants $t_{ab}$ and $\Gamma_{s}$. In fact, by changing the  coupling constants, it is possible to make some peaks of the LDOS so small that the current can be sensitive to the reduction of the channels even in polarization values close to unity and the NDC can be recovered. In fact, in the corresponding cases of the figure \ref{LDOSP080}, the LDOS displays some peaks almost totally suppressed. An example is shown in the figure \ref{LDOSP080} where is plotted the LDOS for $P=0.80$ and $eV_{a}=eV_{b}=-0.13$.  When we increase the applied bias there is a suppression of the first and third peaks localized at $\omega=-0.54$ and $\omega=0.23$, respectively. The suppression is almost complete for the first peak of the LDOS-A and for the third peak in the LDOS-B curve for $eV=0.57$ and $eV=0.83$. Since the Andreev reflection requires a symmetric pair of channels in order to conduct, the process is dominated by those suppressed peaks allowing to observe the NDC in cases with high values of $P$.

\section{Conclusion}

In this work, we have studied the effects of the interdot and intradot interactions on the transport
properties of double quantum-dot system coupled to a ferromagnet and a superconductor. Energy parameters
of the theory are limited by the size of the superconductor gap. This way, the conduction through the device
is controlled by Andreev scattering processes. In the first part of the paper, the role of the coupling between
dots and the coupling of dots with the electrodes was elucidated. Next, we study the effects of electronic
correlations at the dots, within a mean field approximation. For both interactions, inter and intra-dot
correlations, we found regions of negative differential conductance (NDC).
Correlations tend to localized the electrons at the double-dot system,
changing the LDOS at the dots by suppressing some peaks and shifting their positions, leading to an
asymmetric pattern for the LDOS.

This asymmetry reduces the number of available states to conduct through the
Andreev reflection mechanism. The above phenomena, combined with the transmittance reduction with
the applied bias, produce the NDC effect for some regions of the voltage bias. By applying a gate voltage,
one can tune the effect and change the bias region where NDC appears.
Such kind of devices, as the one considered here, are on the verge of being produced by present technology, and our theoretical study may be useful to control the current in practical applications. $\mathcal{K}$ and $\mathcal{U}$ are intrinsic parameters, which are sample dependent. However, their effect can be monitored by the gate voltages, as
shown in this contribution. With the addition of a second ferromagnetic electrode, one may open the possibilities
of crossed Andreev reflections and control of the current by the relative polarization directions of the two
ferromagnets.

The results presented in this work were obtained from a mean field theory and have to be interpreted with attention. Indeed, one could argue that the NDC could be washed out by fluctuations. However, the results shown in the figures \ref{varpolarizacao} and \ref{LDOSP080} were obtained for high values of polarization and nonzero gate voltages where the mean field approximation works well since the high values of the polarization reduce the spin fluctuations. Therefore, we believe that the NDC is a real effect and not only a result of the approximation.  However, the exact extension of the validity of the approximation used in this work can be addressed only by experiments.

\begin{acknowledgments}
The authors acknowledge partial support from the Brazilian agency \emph{Conselho Nacional de Desenvolvimento Científico e Tecnológico} (CNPq).
\end{acknowledgments}

\bibliography{biblio}
\end{document}